
\input smgdefs
\twelvepoint\doublespace\parindent=0pt
\centerline{\bf Calculation of electronic properties of amorphous alloys}
\smallskip
\leftline{J.C. Swihart$^a$,
D.M.C. Nicholson$^b$*, G.M. Stocks$^b$*, Y. Wang$^b$*,
W.A. Shelton$^b$*,}
\leftline{ and H.~Yang$^a$}
\leftline{$^a$Physics Dept.,Indiana Univ., Bloomington, IN 47405, USA}
\leftline{$^b$MS 6114 Bldg 4500S, Oak Ridge National Lab.,
Oak Ridge, TN 37831, USA}

\bigskip
\bigskip

\leftline{\bf Abstract}

We describe the application of the locally-self-consistent-multiple-scattering
(LSMS)[1] method to amorphous alloys. The LSMS algorithm is optimized
for the Intel
XP/S-150, a multiple-instruction-multiple-data parallel computer with
1024 nodes and 2 compute processors per node. The electron density at each
site is determined by
solving the multiple scattering equation for atoms within a
specified distance of the atom under consideration.
Because this method is carried out in real space it is ideal for treating
amorphous alloys.  We have adapted the code to the calculation
of the electronic properties of amorphous alloys.  In these
calculations we determine the potentials in the atomic sphere
approximation self consistently at each site, unlike
previous calculations[2]
where we determined the potentials self consistently at
an average site.  With these self-consistent potentials,
we then calculate electronic properties of various amorphous
alloy systems.  We present calculated total electronic
densities of states for amorphous Ni$_{80}$P$_{20}$ and
Ni$_{40}$Pd$_{40}$P$_{20}$ with 300 atoms in a supercell.

\bigskip
\bigskip

*This work was supported in part by the ORNL Partnership in
Computational Sciences (PICS) Program, supported by the
Department of Energy's Mathematical Information,
and Computational Science (MICS) Division of the Office of Computational
Technology Research.  Additional support was provided by the Division
of Materials Science, Office of Energy Research, U.S. DOE
under contract No. DEAC05-84OR21400 with Lockheed Martin
Energy Systems.

\vfill
\break

{\bf 1. Introduction}

\parindent=20pt

In the past[2] we have carried out calculations of the electronic
properties of various concentrations of NiP amorphous alloys
in the Local Density Approximation (LDA) in which the atoms
were arranged in a supercell and relaxed using Weber-Stillinger
potentials.  The electronic wave functions were determined
by a linearized Korringa-Kohn-Rostocker (KKR) band-structure
calculation within the LDA.  Spherical muffin-tin
potentials were used which, for a particular concentration,
were determined self consistently
from a separate KKR Coherent-Potential-Approximation (CPA)
calculation
on a random alloy in which the atoms were placed at random
on an fcc lattice.  Thus the potentials were not determined
self consistently for the particular arrangement of the atoms
in the supercell for the amorphous alloy.

The results of these calculations produced densities of states
and electrical resistivities due to disorder scattering in quite
good agreement with experimental results, although the variation
of the resistivities with concentration appeared to be somewhat
less than experiment.  We also calculated the thermopower, and we
found a variation with concentration which was in reasonable
agreement with experiment.  However, we did not find a change
in sign of the thermopower with concentration that was found
experimentally.

The LSMS is a real-space, multiple-scattering method for solving the Kohn-Sham
equations that is well-suited for massively parallel computers.
We have made the relatively minor changes in the
code needed to apply this method to supercell models of amorphous
alloys.  The advantages of using this method over our earlier
procedures are several.  First of all, with LSMS we can now determine
the potential self consistently at each site in the supercell.
Secondly, we can go to a larger number of atoms in the supercell.
In our previous serial calculations on a Cray YMP, we were
limited to about
340 atoms in the cell.  With our LSMS calculations on the Intel
Paragon XP/S-150 on which we assign one atom to a node, we can have as
many as 1024 atoms in the cell.  Finally, with our previous serial
calculations on the YMP with 340 atoms, because of memory limitations,
we had to confine our angular momentum decomposition sums to
an $l_{max}=2$.  With LSMS on the XPS150 we can handle up
to $l_{max}=3$ or 4.

In the next section we discuss briefly the LSMS method.  In Sect.~3,
we discuss our calculational procedures.  In Sect.~4, we present
our results for the calculated density of states of amorphous NiP and
amorphous NiPdP alloys.  In Sect.~5 we discuss our results, and in Sect.~6
we give our conclusions.

\leftline{\bf 2. Locally-self-consistent-multiple-scattering method}

The LSMS method[1] is a single-electron,
real-space, multiple-scattering approach based on LDA.  This method is
ideally suited for calculating the electronic properties of a large
number N of atoms (N up to the number of nodes which is currently 1024 on the
XP/S-150) in a supercell by massively parallel
supercomputers because it exhibits linear (O(N)) scaling.  At the heart
of the LSMS method is the observation that a good approximation to the
electron density $\rho^i({\bf r})$ on atom site $i$ can be obtained by
considering only the electronic multiple-scattering processes in a limited
spatial region about the site $i$.  The cluster of M atoms inside this
region is referred to as the Local Interaction Zone (LIZ) for atom $i$.
Each atom is at the center of its own unique LIZ.  In solving the
Schr\"odinger equation
to obtain $\rho^i({\bf r})$ based at site $i$, one replaces the effect of
all of the atoms outside the LIZ of atom $i$ by a constant potential.
The potential that is to be used in the next Self-Consistent Field
(SCF) iteration is calculated by solving Poisson's equation for an
electron density made up of all of the single-site densities
$\rho^i$. A real-space multiple scattering  approach is also  being used by
Arnold and Solberg[3]

In our calculations, we assigned each atom in the supercell to its
won node on the Intel Paragon XPS150 parallel computer at Oak Ridge
National Laboratory.  Starting with an initial assumed potential we
calculate the scattering path matrix $\tau^i_{jk}(\epsilon)$ at
site $i$ with $j,k=1,2,...,M$.  The scattering path matrix is the
inverse of the real space KKR matrix
$t^{-1}(\epsilon)-g(\epsilon)$ where $t$ is the single-site t-matrix,
$g$ is the real space structure constant matrix,
and $\epsilon$ is the electron energy.  With a maximum angular momentum
of $l_{max}=3$, the KKR matrix that must be inverted on each
node is of dimension 16 times M.  The Green's function, and from this
the charge density at site $i$, can be determined from the scattering
path matrix.  With this we then obtain a new potential for the next
iteration.  These iterations are carried to convergence at which
point we have a self-consistent electron potential at each site.

\leftline{\bf 3. Calculational procedures}

For our preliminary calculations using LSMS to determine electronic
properties of amorphous alloys, we have examined amorphous
Ni$_{80}$P$_{20}$ and Ni$_{40}$Pd$_{40}$P$_{20}$.  These are interesting
systems to compare because, although both can be produced as amorphous
systems, the NiPdP system can be made in the amorphous form with
slow cooling from the melt[4] whereas NiP alloys become amorphous only
with very rapid cooling from the melt.  The NiPdP alloys can thus
be produced as amorphous materials in bulk three-dimensional samples.
Such alloy systems could have important potential applications.
Thus it is of interest to determine on a microscopic scale
the differences between these two systems.

We have used the atomic sphere approximation (ASA) with overlapping
spheres to represent our potential.  The ASA sphere volume at a site is taken
to
be equal to that of the Voronii polyhedron surrounding it. Thus  each site has
a
different ASA sphere radius.

We report here on results for
the electronic density of states determined using one sample of
300 atoms for the supercell for each of the two alloys. The sample was
constructed by relaxing  with pair potentials a dense random packing of hard
spheres as described in reference[2]. The sample was shown to agree with
experimental partial pair distribution functions of Ni$_{80}$P$_{20}$.
Unfortunately partial pair distribution functions are not available for
Ni$_{40}$Pd$_{40}$P$_{20}$. Furthermore, the density has not been published.
We therefore took the sample for Ni$_{40}$Pd$_{40}$P$_{20}$ to be identical to
that for Ni$_{80}$P$_{20}$ except that half the Ni atoms were replaced at
random with Pd.

The LIZ was taken to have a radius of 5.1AU which includes the nearest neighbor
shell of atoms. Other calculations indicate that this size LIZ is sufficient to
give electron densities and potentials such that evaluation of the LDA energy
using these electron densities and potentials and evaluating the eigenvalue
sum using a large LIZ radius (approximately 10.0AU) will give total energies
accurate to tenths or hundredths of a mRy. The fact that the energy can
be accurately determined using electron densities calculated with a small
LIZ radius is related to the stationarity of the energy with respect to the
electron density. Unfortunately there is no stationarity principle to insure
the accuracy of the density of states. We anticipate that the density of
states shown here will be very close to densities of states calculated with a
LIZ radius taken to convergence.

\leftline{\bf 4. Results for NiP and NiPdP amorphous alloys}

After iterating the calculations for the self-consistent potential to
convergence for the Ni$_{80}$P$_{20}$ sample of 300 atoms, we can
examine the charge transfer at each Ni and at each P site.  In terms
of the number of electrons transferred to a particular site (a positive
number means that electrons , or negative charge, has been transferred
to that site while a negative number means just the opposite), we find
that the maximum transfer to the 240 Ni sites is 0.574, the minimum is -.096,
and the average transfer is 0.169.  For the 60 P sites, the maximum is
-0.269, the minimum is -0.997, and the average is -.675.

For both the Ni$_{80}$P$_{20}$ and the N$_{40}$Pd$_{40}$P$_{20}$ samples,
we calculated the total electronic density of states for the converged
potentials.  Fig.~1 shows the calculated density of states for
Ni$_{80}$P$_{20}$.  The solid curve is the result of the present
calculation in which the potential is determined self consistently
at each site, while the dotted curve is for a previous calculation [2]
in which the potential was not determined self consistently.

Fig.~2 gives the result of our calculation of the electronic density of
states for our 300 atom sample of Ni$_{40}$Pd$_{40}$P$_{20}$ with the
potential determined self consistently at each site.

\leftline{\bf 5. Discussion}

The amount of charge transferred at each atomic site compared to the
neutral atom depends very much on the particular volume taken around
the atom.  Thus the actual value for the charge transfer has
limited significance.  However, a comparison of the charge transfer
values among all the Ni atoms or among all the P atoms gives an
indication of the variation of local environments among all of the
atoms of a given species in the sample.  The point is that the large
variation that we find for our samples of amorphous Ni$_{80}$P$_{20}$
and Ni$_{40}$Pd$_{40}$P$_{20}$ after the potentials have been
determined self consistently indicate that there are large variations
in the potentials among the atoms of a given species in the sample.

On comparing the electronic density of states of Ni$_{80}$P$_{20}$
calculated here with self-consistent potentials (Fig.~1, solid line)
and an earlier calculation[2] in which there was one Ni potential used
for all the Ni sites and one P potential used for all the P sites
(Fig.~1, dotted line), we note several points of qualitative agreement.
Both
calculations exhibit a large d-band peak of roughly the same width,
and the position of the peak is at about the same energy with respect
to the Fermi level.  Also the density of states at the Fermi level
is approximately the same in the two cases.  This last value has
important consequences in determining the specific heat and the
transport properties.  Our earlier calculation produced quite good
agreement with experiment for these properties. Serious calculation of
the specific heat and transport properties would entail greatly extending
the LIZ. It is likely that such an extension of the LIZ would broaden the
density of states and move the Fermi level up slightly. Therefore at this
point we are not concerned about the difference between the LSMS density of
states at E$_F$ and our earlier work.

There are also quantitative differences between the two calculations.
The d-band density of states is broader, and its peak occurs closer
to the Fermi level for the earlier calculation compared to the present
one.  The density of states at the Fermi level is about 15\% lower in
the present calculation than in the earlier one.  Also the present
calculation produced a second smaller peak at about 0.2 Ry below the
Fermi level while the earlier calculation has only a shoulder in
this energy region.

The calculated electronic density of states that we found for
Ni$_{40}$Pd$_{40}$P$_{20}$, Fig.~2, differs considerably from that
of Ni$_{80}$P$_{20}$, Fig.~1.  As one would expect, because Pd is
a much larger
atom than Ni, the overlap between Pd d-orbitals and the d-orbitals on
neighboring Ni and Pd atoms results in a density of states  with a much
wider d-band peak as shown in Fig.~2. Recall that we have fixed the atomic
positions at the values appropriate for Ni$_{80}$P$_{20}$, with no scaling to
larger volume to accommodate the larger Pd atoms. However the pressure
calculated at this volume is only slightly positive indicating that the
equilibrium volume is not much larger for Ni$_{40}$Pd$_{40}$P$_{20}$ than
for Ni$_{80}$P$_{20}$. Relaxations other than simple scaling will tend to
increase the Pd-Pd and Pd-Ni bond lengths and narrow the density of states.
The density of states at the Fermi level is approximately 25\% smaller
for the Ni$_{40}$Pd$_{40}$P$_{20}$ compared to the Ni$_{80}$P$_{20}$.

\leftline{\bf 6. Conclusions}

We have demonstrated that we can apply the LSMS method to
amorphous alloys represented by a supercell of 300 atoms
and can determine
the potential at each atomic site self consistently
on a massively parallel supercomputer.
We have found that the charge transfer varies appreciably
from site to site for the same species in a given sample
indicating that the use of potentials made self consistent
only at an average site is of limited validity.
We have also demonstrated that we can use the the fully
self-consistent potentials to
calculate electronic properties of these alloys.
We have found the electronic density of states for one
sample with 300 atoms representing amorphous Ni$_{80}$P$_{20}$
and one sample with 300 atoms representing amorphous
Ni$_{40}$Pd$_{40}$P$_{20}$.

For future work, we plan to go to larger samples and to explore
more fully the dependence of our results on both sample size and
the size of the size of the local interaction zone.  We also plan
to extend the calculations to determining transport properties and
the free energy.
\vfill
\break

\vskip 2pt
\bigskip
\bigskip

\leftline{\bf References}
\bigskip
[1]  D.M.C.~Nicholson, G.M.~Stocks, Y.~Wang, W.A.~Shelton, Z.~Szotek,
and W.M.~Temmermann, {\sl Phys.\ Rev.\/} B50 (1994) 14686; Y.~Wang,
G.M.~Stocks,
W.A.~Shelton, D.M.C.~Nicholson, Z.~Szotek, and W.M.~Temmermann,
{\sl Phys.\ Rev.\ Lett.\/} Vol. 75 (1995) 2867.

[2] H.~Yang, J.C.~Swihart, D.M.~Nicholson, and
R.H.~Brown, {\sl Phys.\ Rev.\/}  B47 (1993) 107;
and in {\sl Materials
Theory and Modelling}, ed.~J.~Broughton, P.D.~Bristowe,
and J.M.~Newsam,
{\sl Mat.~Res.~Soc.~Symp.~Proc.}, Vol. 291 (1993) p.~419.

[3] R.~Arnold and H.~Solbrig, {\sl J.~Non-Cryst.~Solids}
189 (1995) 129; and paper PA5, Ninth International Conference
on Liquid and Amorphous Metals, 1995.

[4] A.J.~Drehman and A.L.~Greer, {\sl Acta Metall.}
32 (1984) 323.

\bigskip
\bigskip

\leftline{\bf Figure Captions}
\bigskip

Fig.~1. Calculated electronic density of states (in Ry$^{-1}$
atom$^{-1}$ for both spins) for amorphous
Ni$_{80}$P$_{20}$.  The solid line is for our present calculation
with 300 atoms in a supercell and
using LSMS converged to self-consistent potentials at each site.
The dotted curve is from a previous calculation[1] and represents
an average over several supercells with 160 atoms in each cell.
For the dotted curve, the potentials were self consistent only for
an average site.

\bigskip

Fig.~2. Calculated electronic density of states (in Ry$^{-1}$
atom$^{-1}$ for both spins) for amorphous
Ni$_{40}$Pd$_{40}$P$_{20}$ with 300 atoms in a supercell.
For the calculation we
used LSMS converged to self-consistent potentials at each site.

\bye